# DIFFERENTIALLY PROCESSED OPTIMIZED COLLABORATIVE RICH TEXT EDITOR

Nishtha Jatana[1], Mansehej Singh[2], Charu Gupta[3], Geetika Dhand[4], Shaily Malik[5], Pankaj Dadheech[6], Nagender Aneja[7,8,*], Sandhya Aneja[9]

[1,2,4,5] Maharaja Surajmal Institute of Technology, New Delhi, India
[3] Bhagwan Parshuram Institute of Technology, Delhi, India
[6] Swami Keshvanand Institute of Technology, Management & Gramothan, Jaipur, Rajasthan, India
[7] School of Digital Science, Universiti Brunei Darussalam, Brunei Darussalam
[8] Department of Computer Science, Purdue University, West Lafayette, IN, USA
[9] School of Computer Science and Mathematics, Marist College, Poughkeepsie, NY, USA

[1]nishtha.jatana@gmail.com, [2]mansehej@gmail.com, [3]charugupta@bpitindia.com, [4]geetika.dhand@gmail.com, [5]shaily.singh99@gmail.com, [6]pankajdadheech777@gmail.com, [7]nagender.aneja@ubd.edu.bn, [8]naneja@purdue.edu, [9]sandhya.aneja@marist.edu

**ABSTRACT**
A collaborative real-time text editor is an application that allows multiple users to edit a document simultaneously and merge their contributions automatically. It can be made collaborative by implementing a conflict resolution algorithm either on the client side (in peer-to-peer collaboration) or on the server side (when using web sockets and a central server to monitor state changes). Although web sockets are ideal for real-time text editors, using multiple collaborative editors on one connection can create problems. This is because a single web connection cannot monitor which user is collaborating on which application state, leading to unnecessary network queries and data being delivered to the wrong state. To address this issue, the current solution is to open multiple web socket connections, with one web socket per collaboration application. However, this can add significant overhead proportional to the number of apps utilized. In this study, we demonstrate an algorithm that enables using a single web socket for multiple collaborative applications in a collaborative editor. Our method involves modifying the socket's code to track which application's shared state is being worked on and by whom. This allows for the simultaneous collaboration of multiple states in real-time, with infinite users, without opening a different socket for each application. Our optimized editor showed an efficiency improvement of over 96% in access time duration. This approach can be implemented in other collaborative editors and web applications with similar architecture to improve performance and eliminate issues arising from network overload.

**Keywords:** Collaborative editor; Optimization; Computer-supported Cooperative Work; Conflict Free Replicated Data Types; Rich-text Editor

**Nomenclature:**

| | |
|---|---|
| WYSIWYG | What You See Is What You Get |
| CRDT | Conflict-free Replicated Data Type |
| OT | Operational Transform |
| XML | Extended Markup Language |
| HTML | Hyper Text Markup Language |
|  |  |

# I. INTRODUCTION

Internet-based collaboration allows people to synchronize their ideas and skills to achieve a desired task [1]. The workplaces today are heavily leaning towards collaborative contributions rather than individual work. Collaboration makes it easier to brainstorm ideas to solve an existing problem or deliver the required tasks within the deadline. Working online and the work-from-home culture of today's world demands this collaboration to be made available without the constraints of physical proximity or limited technical tools. It is a necessity for working individuals to be able to collaborate effectively with their colleagues irrespective of their physical geographical locations.

This paper analyzes the task of solving the constraints of collaborative editing on a Rich-Text WYSIWYG (What You See Is What You Get) editor [2]. A Rich-Text Editor allows users to format their text via a toolbar containing an array of icons. Selecting the text and formatting via these icons allows the user to alter the format of their text. One can control various aspects of their text and its font, such as its size, weight, family, style, indentation, and many more, along with several other features. When made collaborative in the form of a web application or software, such an editor provides a platform for multiple users to connect and edit a single document in real-time with the automatic, conflict-free merging of all of their edits. Making a Rich-Text editor collaborative comes with a cost levied upon the editor's performance. A collaborative Rich-Text Editor tends to have technical issues such as lagging and slow performance when the document size is large or when multiple users edit the document in the editor [3].

In this paper, a collaborative Rich-Text Editor application has been made collaborative with the help of a Conflict-free Replicated Data Type (CRDT) algorithm. ProseMirror [4], an open-source and extendable WYSIWYG editor, is used. This was abstracted for VueJS by a layer of TipTap [5]. This editor is used as it has an extendable schema, modular and state management nature. The library is unopinionated, allowing it to experiment with various collaborative schemes. Multiple instances of TipTap editors were connected across a single WebSocket for the demonstration. The proposed work, is editor-independent and can be used with a different choice of editor or application in general, such as collaborative, whiteboarding, and more.

For rendering purposes, Rich-Text Editors map the domain of what you write onto the editor directly to the Document Object Model (DOM). DOM provides an Application Programming Interface (API) to develop HTML and XML documents. DOM nodes provide a representation of the contents of the document. A Rich-Text document typically comprises paragraphs, headings, images, videos, code blocks, and pull quotes. These nodes may have child nodes, further containing rich-text content. The DOM nodes also comprise properties that are very specific to these nodes, which help render the nodes in the editor. The disadvantage, however, is that as the DOM grows huge, as more nodes are added, the DOM may reflow, resulting in the slow performance of the web page due to multiple reflow operations.

In this work, the application divides a single collaborative editor into multiple collaborative editing applications, termed Doclet. This is done one at a time (only one doclet is being used as a collaborative editing application), and the rest of the doclets are being rendered onto the web page as simple HTML parts. This significantly lessens the effective editor's size on the application. Hence, while working on a very large document, only a small part (a single doclet) as the editor is used. This part reflows and repaints upon the changes made, and the rest of the document is rendered simply as an HTML web page. To change which doclet (viz. which part of the document) is to be used as an active collaborative editing application, the user can click on that part, and the position of the cursor on the whole document determines and differentiates between the doclet, which reflows and is an editor instance vs the doclets rendered simply as HTML pages.

A web socket is a communications protocol that offers full-duplex communication channels on one connection. This stateful, bidirectional protocol aids client-server communication, such as in the proposed work. Once the communication link and the connection are established between the client and the server, message exchanges occur in bidirectional mode until the connection persists between the client-server. In the proposed application, multiple doclets are used, which potentially act as an active collaborative editing application. This tends to paint an illusion that there are multiple clients for one server and a single web socket connecting one server to one client. This creates a huge network overhead as many messages are exchanged over the network for communication. In this work, one of the objectives is to solve this network overhead issue.

The Contributions of the proposed work can be summarized as follows:
- This work aims to address the issue of network overhead encountered while using a collaborative editor that uses web sockets for communication between client and server.
- This work proposes an algorithm that enables using a single web socket for multiple collaborative applications in a collaborative editor.
- The proposed methodology involves modifying the socket's code to track which application's shared state is being worked on and by whom by creating what is known as a doclet.
- The proposed methodology allows for the simultaneous collaboration of multiple states in real-time, with an infinite number of users, without the need to open a different socket for each application, thus optimizing the working of the collaborative editor, and thereby reduces the network delay.

*Outline:* Section II describes the architecture of a typical collaborative editor with multiple instances on the same WebSocket and its components. Section III describes the related work. Section IV describes the proposed work. Section V presents the results. Section VI concludes the work.

## II. ARCHITECTURE OF A TYPICAL COLLABORATIVE EDITOR

Vue.js [6] is a JavaScript framework that constructs web user interfaces. The first version of Vue.js was released in 2014 to make single-page web applications. It's a component-based framework, meaning the whole UI is divided into various components in which we pass data arguments. It has built-in directives for model binding, conditional rendering, and rendering a list of items. It is commonly referred to as Vue and pronounced as "view.js" or "view." It provides a special system referred to as a reactivity system in which each component keeps track of its reactive dependencies, as a result of which it knows when and which components need to be re-rendered. Figure 1 depicts the architecture of a typical collaborative editor having multiple instances on the same WebSocket.

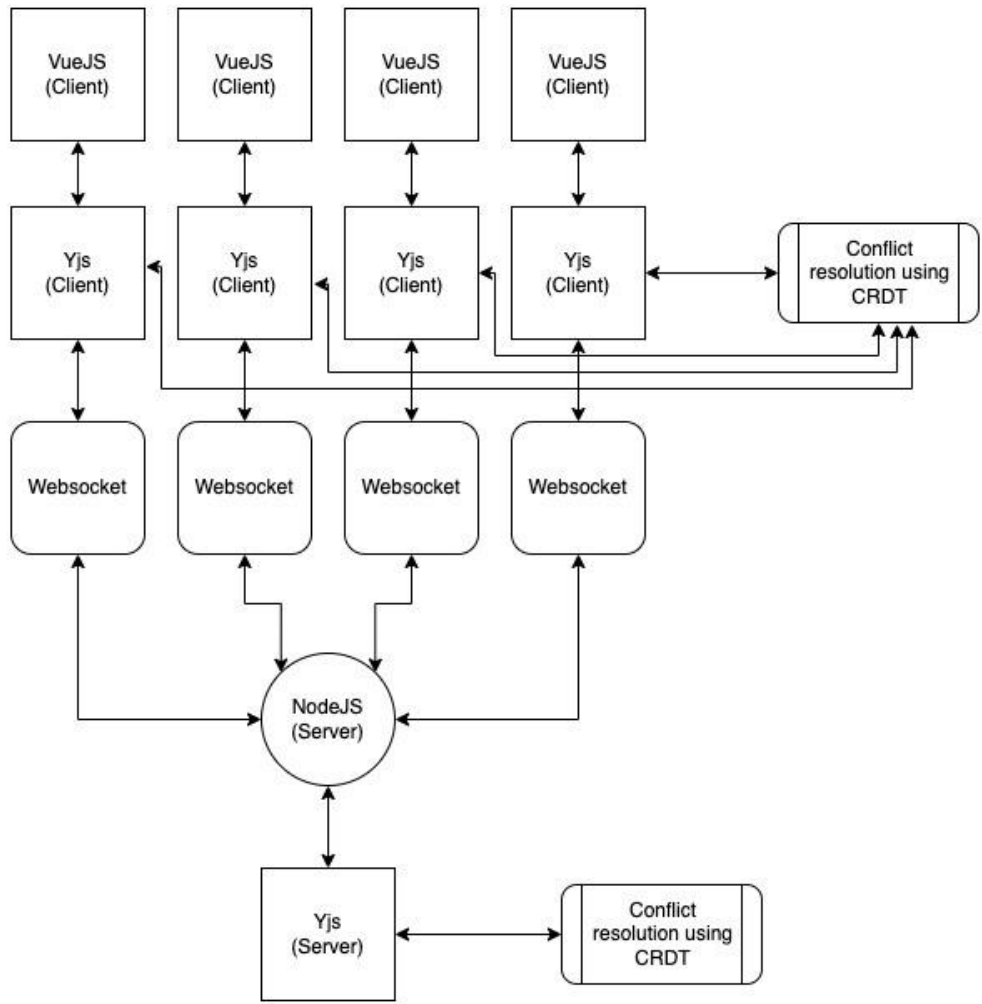

*Figure 1: Architecture of a typical collaborative editor having multiple instances on the same WebSocket*

Yjs* is an open-source CRDT framework built on top of it for building collaborative and real-time sync applications. It uses a "Conflict-Free Replicated" Data Type for syncing automatically. It exposes its internal CRDT model as shared data types that can be manipulated concurrently. In the last decade, real-time editing of documents has become popular for use cases such as writing, drawing, and many more. The available frameworks use an operational transformation approach for real-time data exchange and collaboration. From an engineering perspective, Yjs is easy to use and integrate into web applications. It follows a modular approach that enables any editor to collaborate using any network technology. The main purpose of Yjs is to enable a simple way of building reliable features for collaboration in web applications [7]. Yjs uses P2P to establish reliable real-time sync and collaboration between multiple users so that people can visualize the modifications done by their peers in real-time on a particular document.

*Source: https://github.com/yjs/yjs

On the web, most applications use a request/response model. The client intends to send a request to the server, and the receiving end is a server that responds to it via a suitable response that the client understands based on its configurations. The response may be in HTML, XML, or JSON. When a client requests the server, a pipeline is generated under the hood, through which the data exchange is facilitated by the HTTP protocol. This pipeline is created every time a client requests, irrespective of the previous requests. This eventually results in overhead each time when a request is made. This is where the web sockets [8] emerge. When establishing a socket connection, the server and client can send messages to each other without the overhead of establishing the pipeline, which we discussed earlier. The pipeline is created just once the web socket is established. Now, both parties can exchange data through this web socket connection. When a server or client sends or receives data, the appropriate events are fired through which sent data can be received.

Node.js [9] is a server-side platform built on Google Chrome's JS engine (V8 Engine) for running JavaScript code on a machine. Node.js is a platform built on Chrome's JavaScript for developing scalable network applications and is ideal for data-intensive, real-time applications that span through distributed devices [10]. It's a cross-platform, open-source run-time environment for server-side development. It offers a huge library of various JavaScript modules that further aid the easy development of web applications. There are numerous advantages of using Node.js, including its capability to handle asynchronous and event-driven programs or applications. All APIs in Node.js are synchronous, which signifies that the server never pauses for an API to return data; instead, it simply jumps to the consecutive lines of code to execute them. Being built on V8 Engine, it is very fast in code execution. Although Node.js is single-threaded, it uses a single-threaded model with event looping.

CRDT is capable of managing real-time collaborative text editing using a peer-to-peer framework. No server is used in this algorithm, and if used, it's only for managing connections and coordinating connections among different nodes/users. It even works if clients go offline [11]. It makes changes and synchronizes them as soon as they get the connection. Collaboration algorithms' main goal is consistently broadcasting changes among the different nodes. The result is always the same and consistent on every node, even if multiple users make changes simultaneously. CRDT began emerging around mid-2006 and was defined in 2011. CRDT has two approaches, and both can provide eventual consistency and integrity. They are state-based CRDTs and operations-based CRDTs. These two are theoretically similar, but there are differences in practical implementation. Operation-based CRDTs are called commutative replicated data types (CmRDT), whereas State-based CRDTs are called convergent replicated data types (CvRDT). CmRDT transmits only the update operation, which is commutative and not necessarily idempotent, as compared to the CvRDT, which transmits the full local state to other replicas, where the states are merged by a function that is idempotent, commutative and associative [12]. State-based CRDTs are easier to design and implement than Operation-based CRDTs. Their drawback is that the entire state must be broadcasted, which may be costly [13] [14].

### III. LITERATUR REVIEW

The growth of applications that allow multiple users to edit a document simultaneously and merge their contributions automatically has been observed in the last decade. Jones [2] focused on the changes in modes and uses of writing online. It is further divided into three parts: enabling, documenting, and assessing writing online. The author also gave a list of online editing tools that were available at that time. In [15], it is explained that even though wiki systems are the most popular knowledge-sharing method, they only enable limited cooperative authorship and do not scale effectively. It compares the MediaWiki system with several peer-to-peer approaches for editing wiki pages like MOT2, WOOTO, and ACF. The evaluations are done through qualitative and quantitative metrics. WOOT and ACF are believed to be the most efficient ways of convergence speed regarding rounds and message traffic required. MOT2 and WOOT are adapted for dynamic membership, which is necessary for every P2P network. However, ACF requires a static membership. Lv et al. [16] focused on the CRDT algorithm for integrating string-wise operations for smart and massive-scale collaborations. Under an integrated string-wise structure, this algorithm achieves convergence and retains collaborative users' operation intents. In the age of big data and cloud computing, the CRDT algorithm is being used to promote smart and large-scale collaborations. This algorithm can be used in intelligent and large-scale collaborative applications. The suggested approach has been shown to have lower temporal complexity than the state-of-the-art (OT and CRDT) algorithms, outperforming both. The suggested approach can keep collaborative users' operation intents and the shared document consistent. Zhang [17] modified Storch's model by offering a dyadic interaction model that considers learners' contributions to various areas of collaborative writing and identifies different collaboration techniques. Cluster analysis allowed multiple collaboration patterns to emerge from a dataset based on a quantitative study of learners' comparative involvement in important components of a collaborative writing task. Das et al. [18] are focused on helping visually impaired writers with the help of the Google Colab Docs extension that provides audio output to facilitate understanding who is editing what and where in the shared document. The Colab docs extension uses various spoken and non-spoken speech audio cues to support screen reader users in monitoring others' real-time activities, avoiding concurrent edits. Lee et al. [19] intend to show that the HCI community may support more incisive investigations of LMs' generative capacities by curating and analyzing massive interaction datasets. It begins by describing the generating capabilities of language models; the purpose of NLG is to create fluent text in various domains, such as machine translation. summarization, dialogue, and more. In this paper, the model is enhanced through a pre-trained model. The paper explains the processes to create a dataset that includes 1445 writing sessions featuring rich interactions between 63 writers and four instances of GPT-3 to train and evaluate language models. Rault et al. [20] focused on creating an access control policy and a distributed access control mechanism for collaborative applications based on CRDT Google Docs and POSIX file systems, which are supported by the CRDT as an example of distributed applications. It outlines a generic model data and examines different conflict resolution strategies at the data and policy levels. The concepts used are Security and privacy (Access control), Human-centered computing (Synchronous editors), and Asynchronous editors, which are all examples of computing techniques. Finally, the model must be proven efficient enough to replicate two archetypal cases for decentralized applications. The effects of assessment methodologies (product-based vs. process-and-product-based) on learners' performance and collaborative dynamics in web-based CW are investigated by Zhang et al. [21] [22, 23]. A control group of 82 intermediate tertiary students and an experimental group of 82 intermediate tertiary students were developed. Before completing a synchronous CW task, the control group (n = 20 pairs) was given a standard product-based assessment. In contrast, the experimental group (n = 21 pairs) was given an assessment approach that addressed both the collaborative process and the product. Based on assessments of dyad interaction, revision processes, and co-constructed texts, the experimental group produced texts with better fluency, writing quality, and phrasal complexity. They also exhibited a lot more collaboration during the job. The study provides academicians with a practical strategy for evaluating collaborative writing in the digital era and finds insights that underscore the need for process-and-product-based assessment in tackling web-based CW's difficulties.

The literature contains multiple methods of solving the issues arising in the collaborative text editors. However, the solutions discussed above mainly focus on consistently evaluating the textual content. A few approaches proposed above focus on network delay issues but are complex in their mechanisms. In this paper, the proposed approach aims to resolve the network delay issues that occur in collaborative editors that use CRDT as their underlying approach. The proposed algorithm is simple and intuitive, yielding an optimized editor with reduced network delay.

### IV. PROPOSED WORK

The suggested shared-state tracking architecture's implementation paradigm is described in this section. As a result, it reflects the tracking system's design and implementation and the algorithms that control it. This section also presents the proofs of concepts used in this architecture.

#### A. Proposed Architecture Design

The architecture is an approach for building a shared state class that allows us to track several collaborative states that are being collaborated on consecutively in a single web socket connection. Figure 1 illustrates the architecture layers, components, and tools. This model was based on the fact that an Object-Oriented approach allows us to replicate logic across multiple objects while still maintaining a unique identity for each of them.

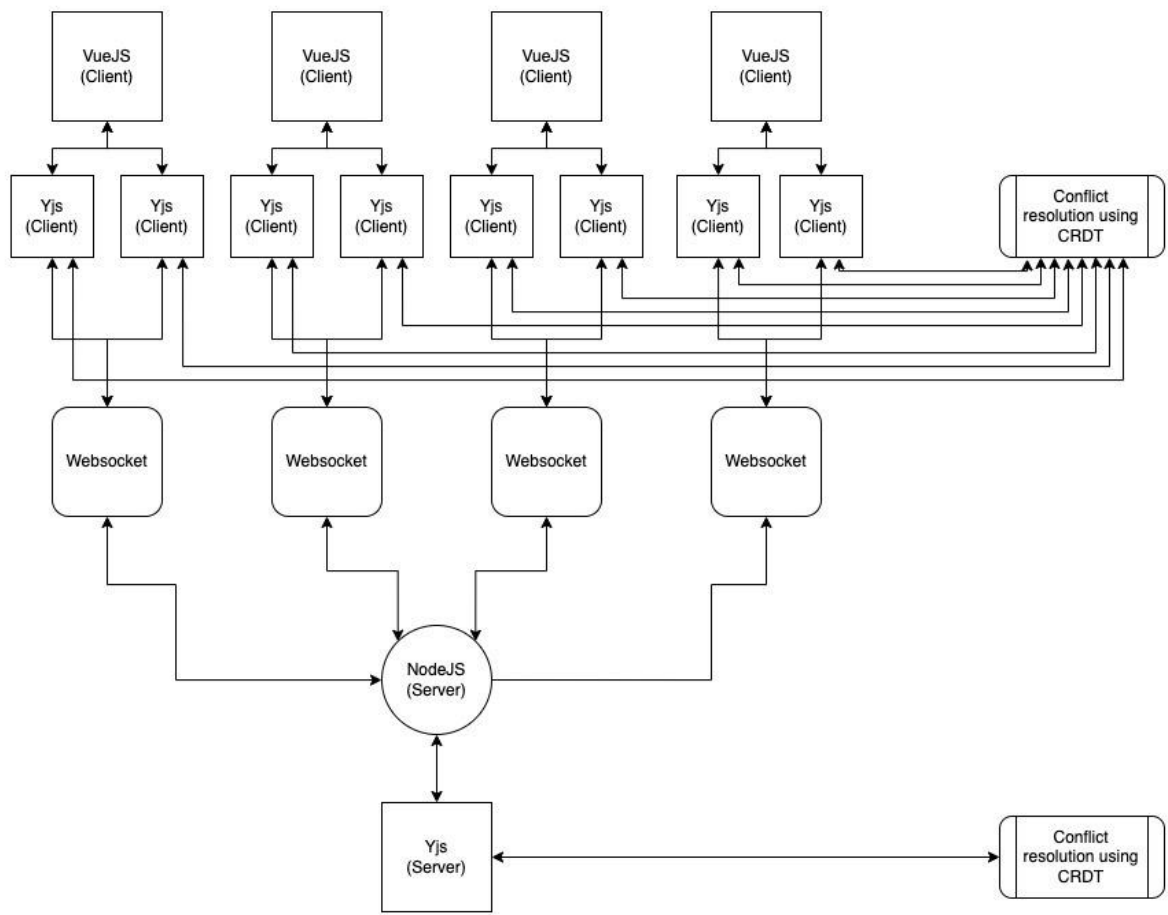

*Figure 2:* PROPOSED ARCHITECTURE DESIGN

Figure 2 represents a three-layer architecture that helps treat different shared states as independent entities while managing them on a single web socket. This is a classic object-oriented approach, where all the shared states share the same class blueprint while maintaining a unique identifier for themselves. In this case, multiple instances of Yjs on the client side are connected onto a single WebSocket per client, a many-to-one connection. Each Yjs instance provides conflict resolution on its shared state using CRDT and passes on the data to a single WebSocket, from where it's then passed on to a server. All collaborative operations are hence performed on the states corresponding to the correct identifier without causing any collisions or merging of data [24-26]. A communication handler layer is set between the web socket and the states, which helps perform state management and comparison operations and passes the messages to the web socket or clients [27-28].

**B. Implementation of Architecture**

We used two TipTap instances on the same connection to implement and demonstrate the architecture.

*Algorithm 1*: **Working of Collaborative Algorithm on a Single Web Socket State**

```
Get editorState, cursorPosition of user
If previousState = null OR previousCursorPosition = null:
        // New instance, or user has just clicked in the editor
        sendToWebsocket(cursorPosition)
If previousState != null:
        onChangeFromWebsocket():
                If previousState != currentState:
                        Update current state with other users' cursors
If previousCursorPosition != null:
        If cursorPosition != previousCursorPosition:
                sendToWebsocket(cursorPosition)
```

As seen in this algorithm, a function tracks the cursor's state and sends messages across the web socket [29-30]. Now, whenever we have multiple cursors and, hence, multiple states, the function gets confused about which state the latest changes are being reflected from [31]. This causes an issue where the cursor jumps across the different state views (ProseMirror instances in our example) and sends infinite network requests to the web socket [32-33].

Ignoring the network's suboptimal behavior also causes the issue that collaborative states' data is transferred to the wrong instances due to the lack of tracking, which change occurs in which state [34-36].

*Algorithm 2*: **Working of Collaborative Algorithm on Multiple Web Socket States**

```
For Each Editor:
        socket = CreateNewWebsocket ()
        Get editorState, cursorPosition of user
        If previousState = null OR previousCursorPosition = null:
                // New instance, or user has just clicked in the editor
                sendToSocket(cursorPosition)
        If previousState != null:
                onChangeFromWebsocket():
                        If previousState != currentState:
                                Update current state with other users' cursors
        If previousCursorPosition != null:
                If cursorPosition != previousCursorPosition:
                        sendToSocket(cursorPosition)
```

This algorithm works fine for a limited number of available states. However, opening multiple web sockets in a single webpage has significant overheads. It's not optimal for more than a few states in a single webpage [37]. Such situations may occur whenever there are multi-content collaborative web pages, wherein collaboration of different types, i.e., text/visual, etc., may be required [38-39].

*Algorithm 3*: Working of Collaborative Algorithm on a Single Web Socket with Multiple States

```
Socket = CreateNewWebsocket()
For Each Editor:
        id = getEditorId()
Get editorState, cursorPosition of user
If previousState = null OR previousCursorPosition = null OR previousID = null:
        // New instance, or user has just clicked in the editor
        sendToSocket(id, cursorPosition)
If previousState != null OR previousID != null:
        onChangeFromWebsocket():
                If previousState != currentState OR previousID != currentID:
                        Update current state with other users' cursors
                If previousCursorPosition != null AND previousID == currentID:
                        If cursorPosition != previousCursorPosition:
                                sendToSocket(id, cursorPosition)
```

Implementing an object-oriented approach, as shown in Algorithm 3, allows us to transform the function discussed in Algorithm 1 into a blueprint that can be replicated across objects while maintaining their unique identity [40-43]. Each unique state view (ProseMirror instance) gets its object bound by this class's behavior, the blueprint [44]. This means that without significantly altering the current state exchange's behavior, we give each state a unique identity and allow tracking using a single web socket connection open [45-50]. Extensive research is underway to enrich internet user experiences for various services available [51-54].

## V. RESULTS

Our proposed application is tested under the following conditions and environment. We took five instances of measurements of the number of WebSocket messages exchanged per second in the idle (non-typing) and busy (typing) state of the optimized editor and the non-optimized editor. After these measurements, we averaged the messages and extrapolated them to get the number of WebSocket message exchanges per 5 seconds.

Results of the time needed for reading using the collaborative editor when in an **Idle (non-typing) state** for a single editor are presented in Table 1. Results of the **Idle (non-typing) state** for two Editors are presented in Table 2. Results of the **Idle (non-typing) state** for four Editors are presented in Table 3. The combined results of the idle (non-typing) state are depicted in Figure 4. This figure depicts the substantial improvement the optimized editor shows in the idle state when the editors working collaboratively are not in the typing state. The overall time gain achieved is 97.18%, which is highly significant; thus, substantial improvement is evident.

*Table 1*: RESULTS: IDLE (NON-TYPING) STATE - Single Editor

| Per Second Measurement | Non-Optimized Editor | Optimized Editor |
|---|---|---|
| Reading #1 | 38 s | 1 s |
| Reading #2 | 29 s | 2 s |
| Reading #3 | 36 s | 0 s |
| Reading #4 | 36 s | 0 s |
| Reading #5 | 33 s | 1 s |
| **Average Per Second Measurement** | **34.4 s** | **0.8 s** |
| **Extrapolated to 5 seconds** | **172 s** | **4 s** |
| **Percentage Decrease (172 -> 4):** | **97.67%** | |

*Table 2*: RESULTS: IDLE (NON-TYPING) STATE – 2 Editor Instances

| Per Second Measurement | Non-Optimized Editor | Optimized Editor |
|---|---|---|
| Reading #1 | 37 s | 1 s |
| Reading #2 | 32 s | 1 s |
| Reading #3 | 37 s | 1 s |
| Reading #4 | 36 s | 0 s |
| Reading #5 | 39 s | 2 s |
| **Average Per Second Measurement** | **36.2 s** | **1 s** |
| **Extrapolated to 5 seconds** | **181 s** | **5 s** |
| **Percentage Decrease (181 -> 5):** | **97.23%** | |

*Table 3:* RESULTS: IDLE (NON-TYPING) STATE – 4 Editor Instances

| Per Second Measurement | Non-Optimized Editor | Optimized Editor |
|---|---|---|
| Reading #1 | 38 | 1 |
| Reading #2 | 50 | 2 |
| Reading #3 | 42 | 1 |
| Reading #4 | 41 | 1 |
| Reading #5 | 39 | 2 |
| **Average Per Second Measurement** | **42** | **1.4** |
| **Extrapolated to 5 seconds** | **210** | **7** |
| **Percentage Decrease (210 -> 7):** | **96.66%** | |

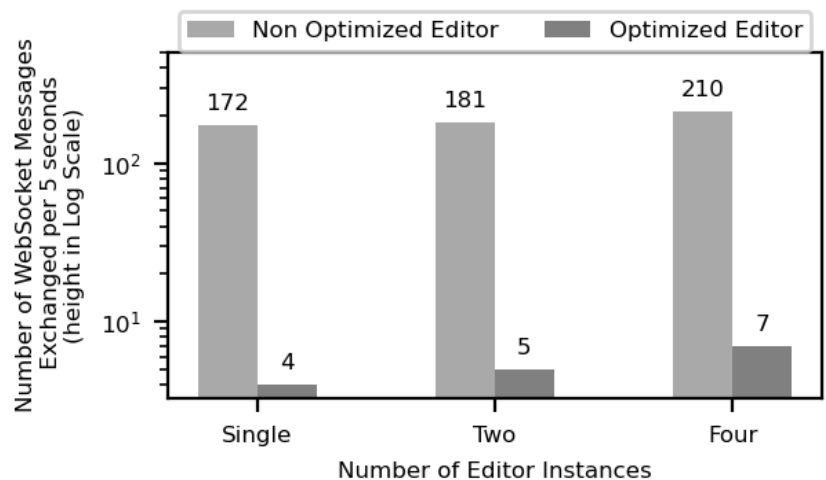

*Figure 3:* RESULTS: IDLE (NON-TYPING) STATE

Results of the time needed for reading using the collaborative editor when in the **active (typing) state** for a single editor are presented in Table 4. The results of the **active (typing) state** for the two Editors are presented in Table 5. Results of the active **(typing) state** for four Editors are presented in Table 6. The results of the idle (non-typing) state are depicted in Figure 3. This figure depicts the substantial improvement in the time needed for reading using the optimized editor in case the editors work collaboratively and edit a particular document. The overall averaged time gain achieved in case of active state is 97.70%, which is again highly significant and thus proves the applicability of our proposed method.

*Table 4*: **RESULTS: ACTIVE (TYPING) STATE - Single Editor**

| Per Second Measurement | Non-Optimized Editor | Optimized Editor |
|---|---|---|
| Reading #1 | 257 | 6 |
| Reading #2 | 312 | 6 |
| Reading #3 | 380 | 5 |
| Reading #4 | 290 | 6 |
| Reading #5 | 392 | 5 |
| **Average Per Second Measurement** | **247.8** | **5.6** |
| **Extrapolated to 5 seconds** | **1239** | **28** |
| **Percentage Decrease (1239 -> 28):** | colspan 97.74% | |

*Table 5*: **RESULTS: ACTIVE (TYPING) STATE – 2 Editor Instances**

| Per Second Measurement | Non-Optimized Editor | Optimized Editor |
|---|---|---|
| Reading #1 | 290 | 6 |
| Reading #2 | 305 | 9 |
| Reading #3 | 335 | 5 |
| Reading #4 | 316 | 7 |
| Reading #5 | 392 | 8 |
| **Average Per Second Measurement** | **327.6** | **7** |
| **Extrapolated to 5 seconds** | **1638** | **35** |
| **Percentage Decrease (1638 -> 35):** | 97.86% | |

*Table 6*: **RESULTS: ACTIVE (TYPING) STATE - 4 Editor Instances**

| Per Second Measurement | Non-Optimized Editor | Optimized Editor |
|---|---|---|
| Reading #1 | 440 | 8 |
| Reading #2 | 390 | 10 |
| Reading #3 | 396 | 9 |
| Reading #4 | 460 | 14 |
| Reading #5 | 408 | 11 |
| **Average Per Second Measurement** | **418.8** | **10.4** |
| **Extrapolated to 5 seconds** | **2094** | **52** |
| **Percentage Decrease (2094 -> 52):** | 97.51% | |

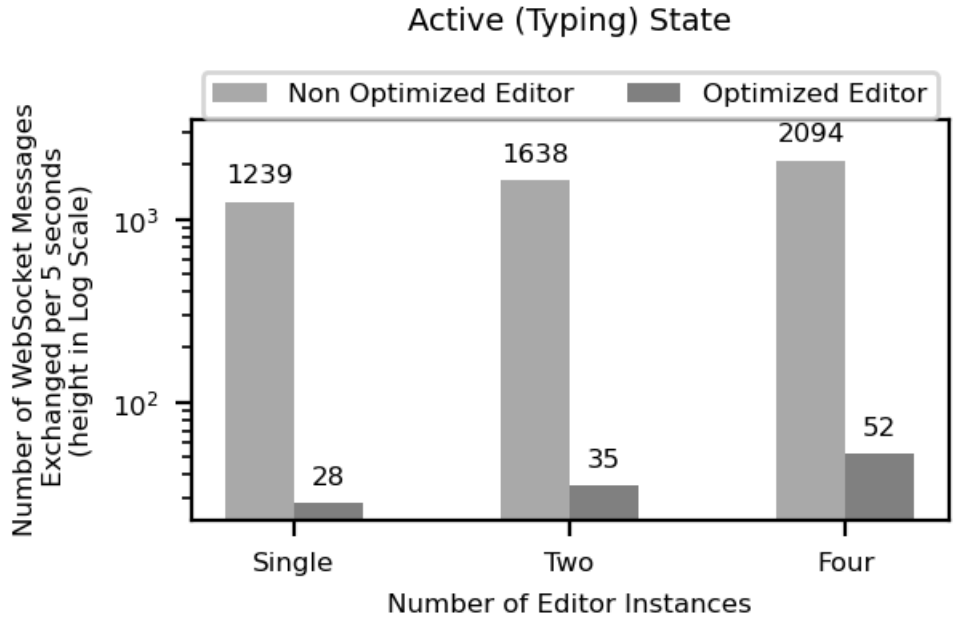

*Figure 4:* RESULTS: ACTIVE (TYPING) STATE

## VI. LIMITATIONS OF THE PROPOSED WORK

The proposed algorithm applies to collaborative editors that use CRDT. The proposed algorithm is not applicable for collaborative editors that use Operational Transform as their underlying approach, as the limitations addressed and associated in this work do not apply to the collaborative editors that use Operational Transform as the underlying approach.

## VII. CONCLUSION

We developed a methodology for a Rich-Text Editor application that allows multiple editor instances to use a single socket for document collaboration. We utilized an Object-Oriented approach and abstracted the messages being passed around the WebSocket into a class. Then, we created a new object with a unique ID mapped to each doclet and deserialized this object to send it as a message to the socket. The overall comparison of the results for Optimized and Non-Optimized editors for Idle and Busy states shows a reading time reduction of 97.67% and 97.74%, respectively (for a single editor instance), in the number of WebSocket messages exchanged. This reduction showcases the substantial effectiveness of the optimization done. Similarly, we found a reduction of 97.23% and 97.86% (for two editor instances) and a reduction of 96.66% and 97.51% (for four editor instances), respectively, for Optimized and Non-Optimized editors. This optimization allows for the effective use of the editor without network overhead obstacles. The proposed approach can be applied to web applications with similar architecture to improve performance and eliminate network overload issues.

## VIII. FUTURE WORK

The rich text collaborative editor that has been optimized by reducing the network delay can further be enriched by the incorporation of the following features:

- Integration of a collaborative whiteboard with the editor using canvas.js that uses socket to exchange information and resolve conflicts as per the CRDT algorithm.

- Integration of Web real-time communication to provide video conferencing amongst the participants that can further provide a smooth and real time collaborative experience.

**Data Availability:** The data are not publicly available

**Conflict of Interest:** The authors declare that they have no conflict of interest.